\documentclass{jpsj2}
\def\runauthor{Online-Journal Subcommittee}

\title{%
Quantum key distribution using vacuum-one-photon qubits: maximum number of transferable bits per particle
}

\author{%
Su-Yong Lee$^{1,}$\thanks{E-mail address: paper1224@kaist.ac.kr},
Se-Wan Ji$^1$, Hai-Woong Lee$^1$, Jae-Weon Lee$^{2,4}$ and J\'{a}nos A. Bergou$^3$
}

\inst{%
$^1$Department of Physics, Korea Advanced Institute of
Science and Technology, Daejeon 305-701, Korea\\
$^2$School of Computational Sciences, Korea
Institute for Advanced Study, 207-43 Cheongryangri-dong, Dongdaemun-gu, Seoul 130-012, Korea\\
$^3$Department of Physics and Astronomy, Hunter College of
The City University of New York, 695 Park Avenue, New
York, NY 10065\\
$^4$Department of Energy Resources Development, Jungwon University, 5 dongburi,
Goesan-eup, Goesan-gun, Chungbuk 367-805, Korea
}

\recdate{\today}

\abst{%
Quantum key distribution schemes which employ encoding on
vacuum-one-photon qubits are capable of transferring more
information bits per particle than the standard schemes employing
polarization or phase coding. We calculate the maximum number of
classical bits per particle that can be securely transferred when
the key distribution is performed with the BB84 and B92 protocols,
respectively, using the vacuum-one-photon qubits. In particular, 
we show that for a generalized B92 protocol with the
vacuum-one-photon qubits, a maximum of two bits per particle can be
securely transferred. We also demonstrate the advantage brought
about by performing a generalized measurement that is optimized for
unambiguous discrimination of the encoded states: the parameter
range where the transfer of two bits per particle can be achieved is
dramatically enhanced as compared to the corresponding parameter
range of projective measurements.
}

\kword{%
Quantum key distribution, Vacuum-One-Photon Qubits
}

\begin{document}
\maketitle

\section{Introduction}

Quantum key distribution (QKD), as first suggested by Bennett
and Brassard \cite{BB84}, provides a way for two parties, known as Alice and
Bob, to share a key with secrecy guaranteed by the laws of quantum
physics. Since its first demonstration in 1992 \cite{Bennett}, QKD has been
extensively studied both theoretically and experimentally to the
point that it now represents probably the most advanced application
of quantum information processing \cite{Gisin}. In most current QKD
implementations, information is encoded either in polarization or in
phase of individual photons. The popularity of the polarization and
phase codings stems from the fact that it is relatively easy and
straightforward to manipulate the polarization and phase of photons
with the already-existing optics technology.

Another possibility for the implementation of QKD protocols is
``photon-number coding'' in which information is encoded on
vacuum-one-photon qubits (VOPQs). By a VOPQ we mean a qubit for
which the two basis states are $|0\rangle$, the vacuum state, and
$|1\rangle$, the one-photon state. In contrast to the QKD schemes
employing polarization or phase coding in which the signal is
carried by individual photons, QKD schemes with VOPQs utilize pulses
in superpositions of the vacuum and one-photon states as the signal
carrier. QKD schemes with VOPQs suffer from an obvious practical
disadvantage that it is generally difficult to generate and detect
such superpositions. Nevertheless, recent theoretical and
experimental progress in the generation and detection of arbitrary
superpositions of the vacuum and one-photon states using techniques
of linear optics \cite{Lund,Pegg,
Dakna,D'Ariano,Paris,Resch,Lvovsky} and cavity QED
\cite{Davidovich,Moussa,Freyberger}, respectively, appears to open
up avenues for quantum information processing with VOPQs. In
particular, quantum teleportation with VOPQs has been studied
theoretically \cite{Lee1,Lee2,Villas-Boas,Koniorczyk,Babichev} and
demonstrated experimentally \cite{Lombardi,Giacomini}. QKD schemes
based on vacuum-one-photon entangled states (single-photon entangled
state \cite{Lee1,Enk}) have also been considered very recently
\cite{Lee3,Giorgi}. Another practical disadvantage of QKD schemes
with VOPQs is that photon losses which are inevitable lead not just
to reduced key rates but also to errors caused by misidentification
of the one-photon state with the vacuum state. This limits the
photon loss rate and consequently the distance over which the key
can be distributed.

In this work we study the BB84 \cite{BB84} and B92 \cite{B92} QKD
schemes using vacuum-one-photon qubits, exploiting in particular the
interesting property of the VOPQ, already noted earlier
\cite{Paris}, that it allows for low-energy-expense encoding of
quantum information, as it requires, on average, less than one
photon for each qubit. The number of photons that Alice needs to
send to Bob in order to transfer a given amount of information can
therefore be smaller with VOPQs than with other qubit systems. This
``cost effectiveness'' of the QKD scheme employing VOPQs is a
practically important characteristic that makes it worthwhile to
study.

The paper is organized as follows. Sec. 2 introduces the parameters
that quantify the cost effectiveness of QKD schemes. In Sec. 3 we
first discuss the cost effectiveness of the BB84 protocol and then
in detail that of the B92 protocol. We first deal with the case when
the encoding states are detected using standard quantum measurements
(Projector Valued Measurements, PVMs). The central result of the
paper is that optimal cost effectiveness can be realized in a much
larger range of the parameters if an optimized generalized
measurement (a Positive Operator Valued Measurement, POVM) is
employed for the detection of the encoding states. In Sec. 4 and 5, respectively, 
we discuss effects of photon losses and of eavesdropping attacks
on QKD schemes using VOPQs. Finally, Sec. 6 presents discussion 
and summary of our main findings.

\section{Effectiveness parameters}

The problem of the cost effectiveness of a QKD scheme that we
deal with here is closely related to one of the most fundamental
issues in information theory, namely, how efficiently information
can be transmitted from input to output of a communication channel.
The classical theory due to Shannon \cite{Shannon} states that the
amount of information that can be transmitted is limited by the
Shannon entropy, a statistical measure of information per ``letter''
of input. 
In quantum communication, which
deals with quantum channels (quantum systems operating as 
communication channels) transmitting signals that are prepared at the input in
the form of quantum states and measured at the output via quantum
state measurements, it is the von Neumann entropy that limits the
amount of information transmitted \cite{Caves,Yuen}. Thus, when
information is carried by qubits, the maximum information per qubit
that can be transmitted is one bit, which is referred to as the
Holevo limit \cite{Holevo1,Holevo2}. The issue we address in this
paper has to do with the transmission efficiency of a QKD scheme,
i.e., we study the question of what are the maximum bits of
information per qubit and per photon, repectively, that can be
securely transferred from Alice to Bob, while preventing an
eavesdropper, Eve, from acquiring information without being
detected. A similar issue has recently been investigated by Cabello
\cite{Cabello}.

For a quantitative discussion of the cost effectiveness of a
QKD scheme, we first define the ``qubit effectiveness parameter'' H as
\begin{equation}
H = \frac{n_{b}}{n_{q}} \ ,
\label{H}
\end{equation}
where $n_{q}$ is the number of qubits sent from Alice to Bob, and
$n_{b}$ represents the number of classical information bits that are
securely transferred from Alice to Bob. This parameter H, which
gives the ratio of the length of the sifted key to the length of the
raw key, may be considered to measure the degree of effectiveness
with which a given QKD protocol produces a secretly shared key per
qubit sent. The Holevo limit establishes an upper bound
to H, i.e.,
\begin{equation}
H \leq 1 \ .
\label{Hbound}
\end{equation}

Another parameter of interest that we wish to consider is the ``particle effectiveness parameter'' or the ``cost effectiveness parameter'' K defined as
\begin{equation}
K=\frac{n_{b}}{n_{p}} \ ,
\label{K}
\end{equation}
where $n_{p}$ is the number of particles (photons) sent from Alice
to Bob. In standard QKD schemes employing polarization or phase
coding, $n_{p}$ and $n_{q}$ are the same and thus $K=H$. When QKD
schemes with VOPQs are considered, however, $n_{p}\leq n_{q}$ and
thus $K\geq H$. We note that the vacuum plays the role of a part of
the signal in the QKD schemes with VOPQs, and yet it does not cost
to send vacuum. Thus, the parameter K can be regarded as a measure
of the cost effectiveness of the QKD scheme being employed. Since K
is not limited by Holevo's theorem, there is no reason why K cannot
be greater than 1 when a QKD scheme with VOPQs is employed.
This suggests a possibility of transferring more than 1 bit of
classical information per photon sent, which is impossible with
other QKD schemes.

\section{Quantum key distribution with vacuum-one-photon qubits}

In this section we calculate H and K for the two best known QKD
protocols, BB84 \cite{BB84} and B92 \cite{B92}, when VOPQs are used
as well as when polarization qubits are used. 
\subsection{The BB84 protocol}

For the standard BB84 protocol with polarization qubits, it is
obvious that $H=K=\frac{1}{2}$, because Alice's and Bob's bases
coincide half of the times. The only difference that arises when we
consider BB84 with VOPQs instead of the polarization qubits is that
the four states $|0\rangle, |1\rangle,
|+\rangle=\frac{1}{\sqrt{2}}(|0\rangle+|1\rangle)$ and
$|-\rangle=\frac{1}{\sqrt{2}}(|0\rangle-|1\rangle)$ in which the
signals are carried are to be interpreted as the vacuum state,
one-photon state, and the symmetric and antisymmetric superpositions
of the vacuum and one-photon states, respectively. Since the signal
is chosen randomly from the four states, the vacuum and one photon
are represented equally on average, and therefore
$n_{p}=\frac{1}{2}n_{q}$. We thus obtain $H=\frac{1}{2}$ and $K=1$.
The K value for BB84 with VOPQs is twice that for BB84 with
polarization qubits.

\subsection{The B92 protocol}

We next consider the B92 scheme. Let us first describe briefly the
standard B92 protocol. We assume polarization coding for convenience
of discussion, although phase coding was considered in the original
proposal \cite{B92}.

\emph{i. Detection based on projector valued measurements (PVMs, standard quantum measurements)}. In the standard B92 scheme with polarization qubits, Alice sends a
sequence of photons in a polarization state randomly chosen from
$|0\rangle = |\leftrightarrow\rangle$ and
$|+\rangle = \frac{1}{\sqrt{2}}(|\leftrightarrow\rangle + |\updownarrow\rangle)$,
and Bob checks whether a signal is transmitted through his analyzer whose
axis is oriented along the direction randomly chosen from the
directions of $|1\rangle = |\updownarrow\rangle$ and
$|-\rangle = \frac{1}{\sqrt{2}}(|\leftrightarrow\rangle - |\updownarrow\rangle)$.
Only when Bob detects a transmitted signal, which occurs with a
probability $\frac{1}{4}$, can he determine with certainty its
polarization state. Thus, for this standard B92 scheme, we have
$H=K=\frac{1}{4}$.

The security of the B92 scheme rests on the fact that arbitrary two
nonorthogonal states cannot be distinguished perfectly. The two
states that Alice chooses from do not need to be $|0\rangle$ and
$|+\rangle$; one can generalize the above B92 scheme to
let Alice choose randomly from two nonorthogonal states
$|\psi_{0}\rangle=\cos\theta_{0}|0\rangle+e^{i\phi_{0}}\sin\theta_{0}|1\rangle$ and
$|\psi_{1}\rangle=\cos\theta_{1}|0\rangle+e^{i\phi_{1}}\sin\theta_{1}|1\rangle$ $(\langle\psi_{0}|\psi_{1}\rangle\neq0)$. Bob then performs his
measurements choosing the axis randomly from the directions of
$|\psi_{0}^{\bot}\rangle=\sin\theta_{0}|0\rangle-e^{-i\phi_{0}}\cos\theta_{0}|1\rangle$ and
$|\psi_{1}^{\bot}\rangle=\sin\theta_{1}|0\rangle-e^{-i\phi_{1}}\cos\theta_{1}|1\rangle$. The probability for Bob to obtain a conclusive measurement outcome then becomes
$\frac{1}{2}(1-|\langle\psi_{0}|\psi_{1}\rangle|^{2})$. For the
generalized B92 scheme, therefore, we obtain
\begin{equation}
H = K = \frac{1}{2}(1-|\langle\psi_{0}|\psi_{1}\rangle|^{2}) \ ,
\label{HKforB92}
\end{equation}
which reduces to $H=K=\frac{1}{4}$ when
$|\psi_{0}\rangle=|0\rangle$ and $|\psi_{1}\rangle=|+\rangle$, as expected.

The generalized B92 scheme can be adopted for VOPQs
in a straightforward way. Here, of course, the two
states $|\psi_{0}\rangle$ and $|\psi_{1}\rangle$ that Alice chooses
from are superpositions of the vacuum and one-photon states. Bob
then needs to perform projection measurements by applying a
projection operator randomly chosen from
$P_{0}^{\bot}=1-|\psi_{0}\rangle\langle\psi_{0}|$ and
$P_{1}^{\bot}=1-|\psi_{1}\rangle\langle\psi_{1}|$. As before with
the polarization qubits, H is given by $H=\frac{1}{2}(1-|\langle\psi_{0}|\psi_{1}\rangle|^{2})$. On the other hand, the average number of photons contained in each qubit is
less than one, and thus K is not equal to H and is given by
\begin{eqnarray}
&&K=\frac{1-|\langle\psi_{0}|\psi_{1}\rangle|^{2}}{\sin^{2}\theta_{0}+\sin^{2}\theta_{1}} \nonumber \\
&=&\frac{1-|\cos\theta_{0}\cos\theta_{1}+e^{-i(\phi_{0}-\phi_{1})}\sin\theta_{0}\sin\theta_{1}|^2}{\sin^{2}\theta_{0}+\sin^{2}\theta_{1}} \ .
\label{KforB92}
\end{eqnarray}

\begin{figure}[ht]
\begin{center}
\includegraphics{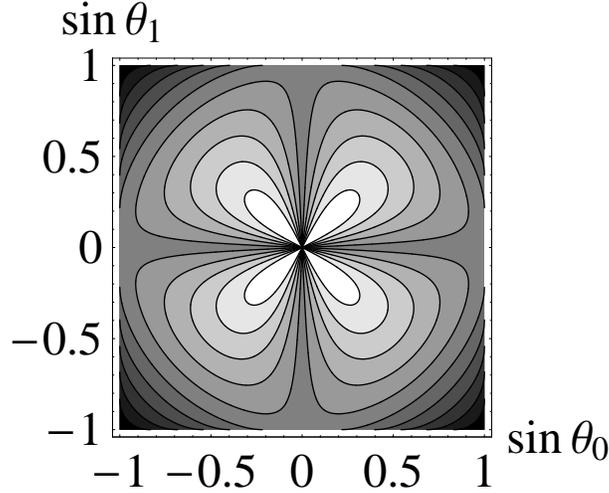} 
\caption{Contour plot of $K_{max}^{PVM}$ as a
function of $\sin\theta_{0}$ and $\sin\theta_{1}$. The magnitude is
represented by brightness from the maximum value of 2 (white) to the
minimum value of 0 (black).} \label{Fig.1}
\end{center}
\end{figure}

It can be seen from Eq. (\ref{KforB92}) that, for given  $\theta_{0}$ and $\theta_{1}$, K takes on its maximum value when $\phi_{0}-\phi_{1}=0$ if $\cos\theta_{0}\cos\theta_{1}$ and
$\sin\theta_{0}\sin\theta_{1}$ are of opposite signs and when
$\phi_{0}-\phi_{1}=\pm\pi $ if $\cos\theta_{0}\cos\theta_{1}$ and
$\sin\theta_{0}\sin\theta_{1}$ are of the same sign. One can thus write $K(\theta_{0},\theta_{1},\phi_{0},\phi_{1})\leq
K_{max}(\theta_{0},\theta_{1}) \equiv K_{max}^{PVM}$, where
\begin{eqnarray}
K_{max}^{PVM} = \frac{1-(|\cos\theta_{0}\cos\theta_{1}|-|\sin\theta_{0}\sin\theta_{1}|)^2}{\sin^{2}\theta_{0}+\sin^{2}\theta_{1}} \ ,
\label{Kmax}
\end{eqnarray}
and the superscript PVM refers to projector valued measurements. In Fig. \ref{Fig.1} we plot $K_{max}^{PVM}$ as a function of $\theta_{0}$ and $\theta_{1}$. $K_{max}^{PVM}$ approaches its maximum value of 2 when both $\theta_{0}$ and $\theta_{1}$ approach 0.

$K_{max}^{PVM}$ takes a simple form if we consider the special case
of $\theta_{0}=-\theta_{1}\equiv\theta$ and $\phi_{0}= \phi_{1}=0$,
i.e., the case where
$|\psi_{0}\rangle=\cos\theta|0\rangle+\sin\theta|1\rangle$ and
$|\psi_{1}\rangle=\cos\theta|0\rangle-\sin\theta|1\rangle$. A
straightforward calculation yields
\begin{equation}
K=K_{max}(\theta)=2\cos^2\theta \ . \label{Kmax2}
\end{equation}
Thus, along the line $\theta_{0}=\pm\theta_{1}$, $K\equiv K_{max}$
varies as $2\cos^2\theta$. For $0 \leq |\theta_{0}| \leq
\frac{\pi}{4}$, the number of bits transferred per photon is between
1 and 2, with the maximum value of 2 obtained for $\theta_{0}=0$.

\emph{ii. Detection based on generalized measurements (Positive Operator Valued Measurements, POVMs)}. To close this section we remark that $K_{max} = 2$ can be reached in
a larger region of the parameters than the one given in
Fig. \ref{Fig.1}. For this, one needs to perform a generalized
measurement (Positive Operator Valued Measure, POVM), instead of the
projective von Neumann measurement leading to Eq. (\ref{HKforB92}).
The POVM that one needs to perform is the one that corresponds to
optimal unambiguous state discrimination \cite{BHH}.
Without going into details we just recall that the corresponding
formulas for this case can be obtained if
$\frac{1}{2}(1-|\langle\psi_{0}|\psi_{1}\rangle|^{2})$ in Eqs.
(\ref{HKforB92}) and (\ref{KforB92}) is replaced by
$(1-|\langle\psi_{0}|\psi_{1}\rangle|)$. This, in turn, yields, in
lieu of (\ref{Kmax}), $K_{max}^{POVM}$ as
\begin{eqnarray}
K_{max}^{POVM}
=2 \frac{1-\mid |\cos\theta_{0}\cos\theta_{1}|-|\sin\theta_{0}\sin\theta_{1}|\mid }{\sin^{2}\theta_{0}+\sin^{2}\theta_{1}} \ .
\label{KmaxPOVM}
\end{eqnarray}

\begin{figure}[ht]
\begin{center}
\includegraphics{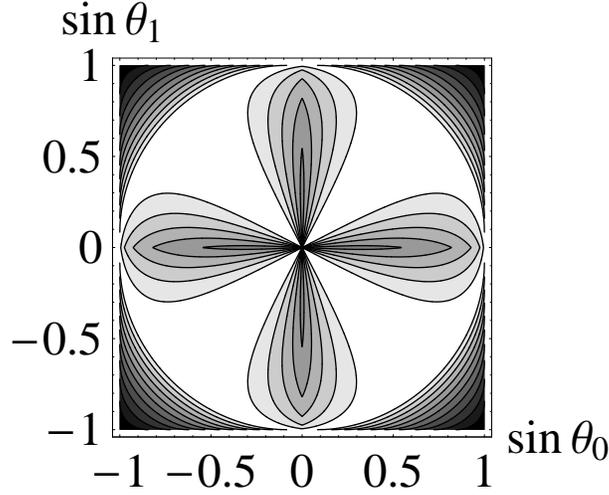} 
\caption{Contour plot of $K_{max}^{POVM}$ as
a function of $\sin\theta_{0}$ and $\sin\theta_{1}$. The magnitude
is represented by brightness from the maximum value of 2 (white) to
the minimum value of 0 (black). A comparison to Fig. 1 reveals the
advantage of a POVM: the parameter range where the maximum value of
2 can be achieved is enhanced.} \label{Fig.2}
\end{center}
\end{figure}

In Fig. \ref{Fig.2} we plot $K_{max}^{POVM}$ as a function of $\theta_{0}$
and $\theta_{1}$. We see that $K_{max}^{POVM}$ reaches its maximum value of 2 for a much larger range of $\theta_{0}$ and $\theta_{1}$ than in Fig. \ref{Fig.1}.

A comparison of Figs. 1 and 2 reveals the advantage brought about by an optimized measurement. While the optimal POVM does not increase the value of $K_{max}$ above 2, it will increase the parameter range where the optimal value can be achieved.

For the special case of $\theta_{0} = -\theta_{1}\equiv\theta$ and
$\phi_{0} = \phi_{1}=0$, considered at the end on the subsection on
PVM detection, i.e., for the case where
$|\psi_{0}\rangle=\cos\theta|0\rangle+\sin\theta|1\rangle$ and
$|\psi_{1}\rangle=\cos\theta|0\rangle-\sin\theta|1\rangle$,
$K_{max}^{POVM} = 2$ along the line $-\frac{\pi}{4} \leq
\theta_{0}=\pm\theta_{1} \leq \frac{\pi}{4}$. Thus, the number of
bits transferred per photon is 2 for a large range of the
parameters, using the optimized measurement.

\section{Effect of photon losses}
Photon losses can be particularly damaging to QKD schemes using
VOPQs, because they induce errors caused by misidentification of the
single photon state $|1\rangle$ with the vacuum $|0\rangle$. In this
section we study effects of the photon losses on  our generalized
B92 scheme using VOPQs.

The effect of the loss of a photon is represented by amplitude
damping which transforms the signal in state
$|\psi_j\rangle=\cos\theta_j|0\rangle+e^{i\phi_j}\sin\theta_j|1\rangle
(j=0,1)$ into a mixed state described by the density matrix
\begin{eqnarray}
\rho^{\prime}_j=\left(\begin{array}{cc}
\cos^2\theta_j+\gamma\sin^2\theta_j &
\sqrt{1-\gamma}e^{-i\phi_j}\cos\theta_j\sin\theta_j\\
\sqrt{1-\gamma}e^{i\phi_j}\cos\theta_j\sin\theta_j &
(1-\gamma)\sin^2\theta_j
\end{array}\right),\nonumber\\
\end{eqnarray}
where $\gamma$ is the probability of losing a photon. Considering
only the photon losses during the fiber transmission, we express
$\gamma$ as
\begin{eqnarray}
\gamma=1-10^{-\alpha l/10},
\end{eqnarray}
where $\alpha$ is the loss coefficient of the fiber and $l$ the
length of the fiber.

With the possibility of photon losses, Alice and Bob now face the
difficulty that the signal transmission through Bob's analyzer does
not necessarily give the correct identification of the signal state.
In order to ensure that our scheme works properly, we require, 
as a minimum condition, that
the probability for correct identification be greater than the
probability for incorrect identification, i.e., we require
\begin{eqnarray}
\min\{Tr(\rho^{\prime}_0P^{\bot}_1),
Tr(\rho^{\prime}_1P^{\bot}_0)\}>\max\{Tr(\rho^{\prime}_0P^{\bot}_0),
Tr(\rho^{\prime}_1P^{\bot}_1)\},\nonumber\\
\end{eqnarray}
where $\min$ and $\max$, respectively, indicate that the smaller and
greater of the two are to be taken. Since the probability for
incorrect identification increases with $\gamma$, the inequality
(11) sets the upper limit $\gamma_{\max}$ on $\gamma$, which in
turn sets the upper limit $l_{max}$ on the key transfer distance
$l$.

In order to illustrate the effect of photon losses, we consider the
case for which
$|\psi_0\rangle=\cos\theta_0|0\rangle+\sin\theta_0|1\rangle$ and
$|\psi_1\rangle=\cos\theta_1|0\rangle-\sin\theta_1|1\rangle$, and
compute $\gamma_{\max}$ and $l_{\max}$ for a fixed value of
$\cos^2\theta_0=0.95$ and different values of $\cos^2\theta_1$. For
our calculation of $l_{\max}$, we choose $\alpha=0.2dB/km$, which is
an appropriate value for fiber transmission at $1550nm$. The result
of our calculation is shown in Figure 3. We see from Fig.3 that, as
the value of $\cos^2\theta_1$ is moved away from
$\cos^2\theta_0=0.95$ toward a smaller value, $\gamma_{max}$ and
$l_{max}$ decrease and thus the amount of photon losses that can be
tolerated decreases. This can be understood by noting that, as
$\cos^2\theta_1$ is decreased, the signal state $|\psi_1\rangle$
moves toward the one-photon state and has  therefore more to lose
when photon losses occur. In fact, one sees from Eq.(9) that
photon losses increase (decrease) the probability to find the signal
in state $|0\rangle$ ($|1\rangle$) by $\gamma\sin^2\theta$. The
smaller $\cos^2\theta_1$ is, the greater the change in the signal
state due to photon losses is, and thus the greater the probability
for incorrect identification of the state becomes. It is interesting
to note, however, that $\gamma_{max}$ decreases sufficiently slowly
with respect to $\cos^2\theta_1$ that the tolerable amount of photon
losses remains high $(\gamma_{max}\gtrsim 0.8)$, as long as
$\cos^2\theta_1$ is reasonably close to $\cos^2\theta_0$. On the
other hand, the maximum key transfer distance decreases rapidly to
less than $\sim 100km$, as $\cos^2\theta_1$ is moved away from
$\cos^2\theta_0$. We note, however, that, as $\cos^2\theta_1$ is
moved away from $\cos^2\theta_0$, the probability
$\frac{1}{2}(1-\left|\langle \psi_0|\psi_1\rangle\right|^2)$ for Bob
to obtain a conclusive measurement outcome increases. We also remark
that the photon loss probability $\gamma_{max}$ at which the
probability for correct identification of the signal state is equal
to the probability for incorrect identification takes on the same
value regardless of whether Bob performs PVM or POVM. Hence, Fig.3
applies also to the case when POVM is performed.

\begin{figure}[tbp]
\centering
\includegraphics[angle=270,width=0.9\columnwidth]{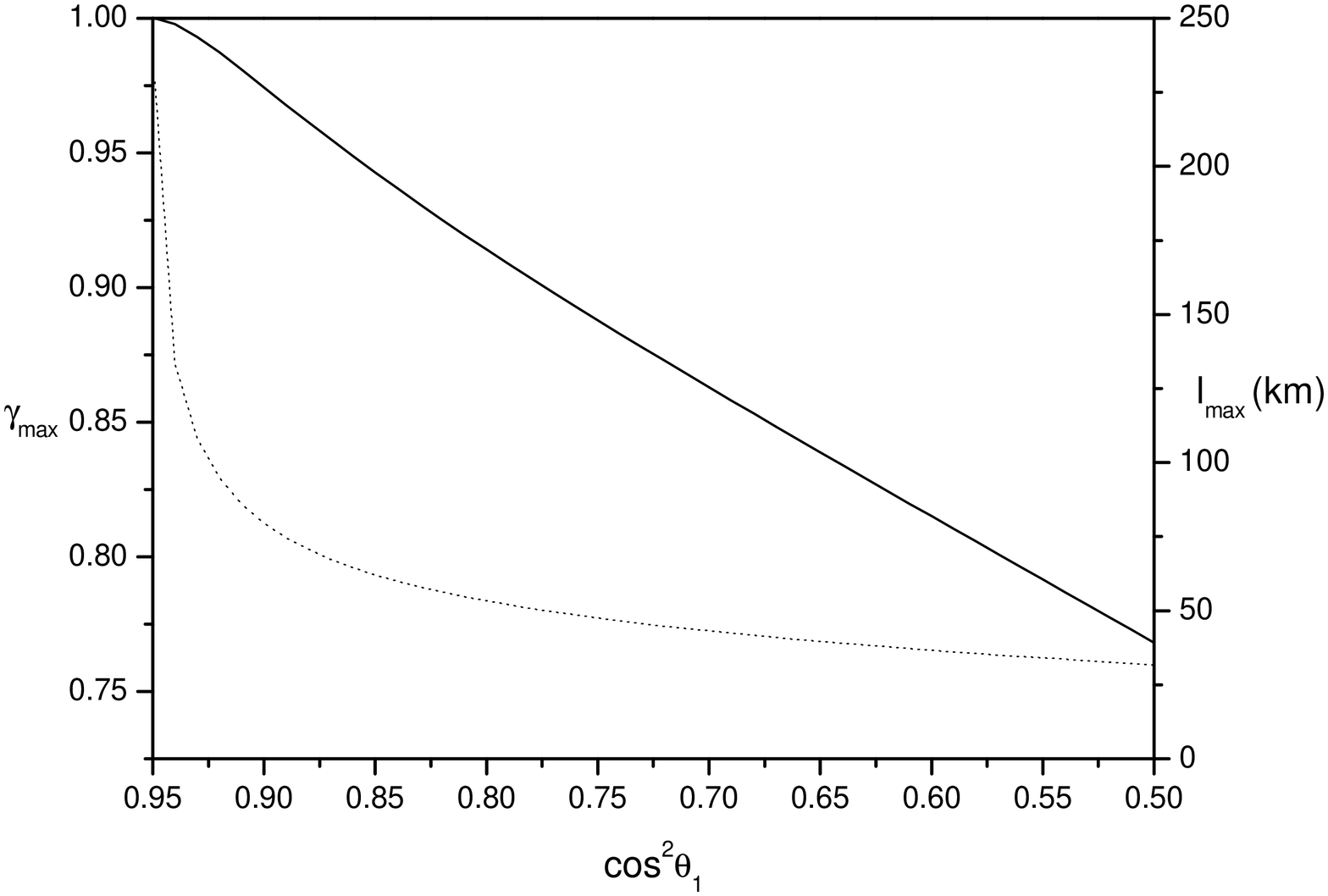}
\caption{$\gamma_{max}$(solid curve) and $l_{max}$(dotted curve) as
a function of $\cos^2\theta_1$. $\cos^2\theta_0=0.95$ and
$\alpha=0.2dB/km$.}
\end{figure}

\begin{figure}[tbp]
\centering
\includegraphics[angle=270,width=0.9\columnwidth]{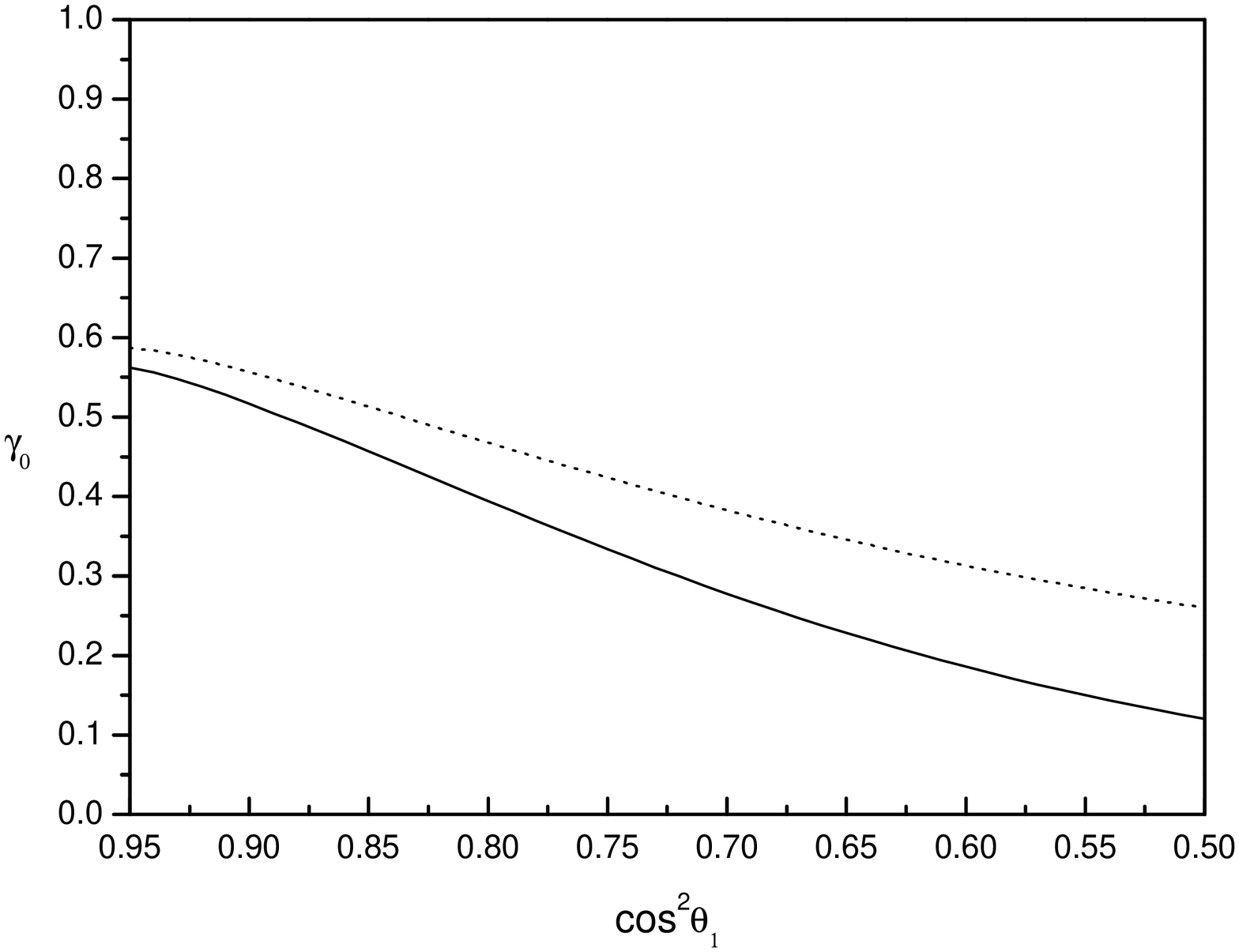}
\caption{$\gamma_0$ as a function of $\cos^2\theta_1$ for the case
of PVM(solid curve) and POVM(dotted curve). $\cos^2\theta_0=0.95$.}
\end{figure}

The photon losses decrease the key rate and thus the parameters H
and K by a factor of $(1-\gamma)$. 
If $\gamma$ is not too large, however, the parameter K can
still be greater than $1$ with our scheme using VOPQs. In Fig.4 we
plot $\gamma_0$ at which $K=1$ as a function of $\cos^2\theta_1$ for
the same case for which Fig.3 is drawn. If $0\leq \gamma <\gamma_0$,
then $K>1$ and thus more than one bit of classical information per
photon can still be transferred in the presence of photon losses. We
see again from Fig.4 the advantage brought about by POVM as compared
with PVM.

\section{Eavesdropping attacks}
In this section we discuss effects of eavesdropping attacks on our key distribution
scheme using VOPQs. For concreteness of discussion we consider the
intercept-resend attack in which Eve intercepts all the signals from Alice
and resends to Bob only those that give her the conclusive measurement outcome.
If Alice and Bob are to detect such eavesdropping attacks, it must be required
that the probability $\gamma$ of losing a photon be less than the probability
$P_{?}$ that Eve's measurement yields an inconclusive outcome, i.e., we have
the requirement on $\gamma$ which reads
\begin{eqnarray}
\gamma < P_?
\end{eqnarray}
Eq.(12) imposes another condition on $\gamma$ in addition to the condition 
derived from Eq.(11).

Let us take, as an example, the case 
$|\psi_0\rangle =\cos\theta |0\rangle+\sin\theta |1\rangle$ and
$|\psi_1\rangle =\cos\theta |0\rangle-\sin\theta |1\rangle$. Assuming
that Eve performs POVM on the signal from Alice, we have 
$P_?=\left| 2\cos^2\theta-1\right|$ and thus Eq.(12) becomes
$\gamma < \left| 2\cos^2\theta-1\right|$. If $\theta$ is close to 0, that is, 
if $|\psi_0\rangle$ and $|\psi_1\rangle$ are both close to vacuum and
thus there is a high probability for Eve to obtain the inconclusive measurement
outcome, then the requirement on $\gamma$ is not so strong. As long as
$\gamma$ is not too large, Alice and Bob can detect eavesdropping attacks,
assuming that Eve blocks all the signals for which she obtains the inconclusive
measurement outcome. On the other hand, if $\theta$ is close to $\frac{\pi}{4}$,
i.e., if $|\psi_0\rangle$ and $|\psi_1\rangle$ are almost orthogonal to each
other and therefore there is only a low probability for the inconclusive measurement
outcome, then the requirement on $\gamma$ is strong. Alice and Bob are forced to use
a low-loss system or are limited to relatively small distances for secure key distribution.

If Alice and Bob find that signal losses are higher than those expected from system
losses $\gamma$, they suspect the presence of Eve and should discard all the
data and restart. They keep the data only when signal losses are within the limit
allowed by system losses only. The parameters H and K are therefore still limited
by system losses, being reduced by a factor of $(1-\gamma)$ compared with
the values in an ideal situation, even if eavesdropping attacks are considered. It
should be mentioned, however, that, in our discussion so far, we have neglected
the reduction of the key length caused by the acts of information reconciliation
and privacy amplification that Alice and Bob need to perform to increase the
accuracy and security of their shared key. The actual values of H and K will
therefore be further reduced. Considering photon losses and effects of
information reconciliation and privacy amplification, the cost effectiveness parameter
K cannot reach its ideal maximum value of 2. It can, however, be still greater than
1, if photon losses are sufficiently small and information reconciliation and privacy 
amplification are efficiently performed. It is possible to securely transfer more than
1 bit of classical information per photon sent using VOPQs.

\section{Discussion and Conclusion}

As described in the previous sections, QKD schemes with VOPQs have
the potential of transferring more than one bit of classical
information per photon. When the B92 protocol is considered, the
number of bits transferred per photon has an upper bound of 2. The
physical reason for the existence of the upper bound on K can be
understood as follows. One might naively expect that K (number of
bits transferred per particle sent) can be made arbitrarily large
simply by choosing the qubit states $|\psi_{0}\rangle$ and
$|\psi_{1}\rangle$ arbitrarily close to vacuum, i.e., by choosing
$\theta_{0}$ and $\theta_{1}$ arbitrarily close to 0, thereby
reducing the number of particles sent from Alice to Bob. This
expectation is, however, inaccurate because, as both $\theta_{0}$
and $\theta_{1}$ approach 0 together, the two states
$|\psi_{0}\rangle$ and $|\psi_{1}\rangle$ approach each other and it
becomes increasingly difficult to distinguish between them, reducing
also the number of bits that can be transferred. The maximum value
of 2, however, is nontrivial, because it does not correspond to the
case when the average photon number per signal is 1/2.

The fact that K can be as large as 2 indicates that two bits
of classical information can be securely transferred by sending just
one photon, if VOPQs are employed. In practice, of course, the
actual value of K should be lower than 2, because a part of the
sifted key is used for checking against eavesdropping and also for
information reconciliation and privacy amplification as described in
Sec. 5.

It should be noted that, as $\theta_{0}$ and $\theta_{1}$ approach
0, the value of K increases, but at the same time the probability
for Bob to obtain a conclusive measurement outcome decreases, which
works to reduce the key rate. For example, when we consider the
simple case in which
$|\psi_{0}\rangle=\cos\theta|0\rangle+\sin\theta|1\rangle$  and
$|\psi_{1}\rangle=\cos\theta|0\rangle-\sin\theta|1\rangle$, the key
rate is given by $(r\sin^{2}2\theta)/2$, where $r$ denotes the raw
key rate, i.e., the number of qubits that Alice sends to Bob in unit
time. Obviously, the key rate approaches 0 as $\theta$ approaches 0.
In practice, therefore, a compromise should be made between the cost
effectiveness which pushes $\theta$ toward 0 and the key rate which
pushes $\theta$ toward $\frac{\pi}{4}$. Clearly, $2\theta = \pi/4$
is a good compromise, in agreement with the original suggestion in
the B92 protocol \cite{B92}, and still comfortably
achieves $K_{max}=2$ if the optimal POVM is employed for the
detection of the coding states.

We mention that the efficiency of a QKD protocol has also been
investigated recently by Cabello \cite{Cabello}. While we consider
here the possibility of going beyond the Holevo limit, he studied
the efficiency of QKD protocols within the Holevo limit. He
considered a parameter E defined as the number of secret bits
transferred per qubit per classical bit of information exchanged
between Alice and Bob through a classical channel. Since the
classical communication between Alice and Bob is generally necessary
in a QKD protocol, we have $E\leq H$. For example, we have
$E=\frac{1}{4}$ while $H=\frac{1}{2}$ for the standard BB84
protocol. Cabello has shown that it is possible to design a protocol
for which Bob's state discrimination succeeds with $100\%$
probability and thus no classical communication is needed. In such a
case, the parameter E is equal to $H$ and can take on its maximum
value $E=1$ in the Holevo limit.

Although QKD schemes with VOPQs have an advantage of being cost
effective, they suffer from a practical difficulty of having to deal
with superpositions of the vacuum and one-photon states. Another
difficulty is with channel losses that are inevitably present. While
channel losses reduce the key rate in the usual QKD schemes
employing polarization or phase coding, they induce errors in QKD
schemes with VOPQs. The requirement on channel losses for a
successful execution of the QKD schemes with VOPQs is more stringent
than that with other types of qubits, as shown in Sec. 4.

In conclusion, we have shown that QKD schemes with VOPQs are capable
of transferring more than one bit of classical information per particle.
When the B92 protocol is chosen, a maximum of two bits per photon can be
securely transferred in an ideal situation with the use of vacuum-one-photon qubits. 
This maximal transfer rate can be reached in a large range of parameters 
if the optimal unambiguous discrimantion strategy is used for the detection 
of the coding states.

\begin{acknowledgments}
S. Y. Lee, S. W. Ji and H. W. Lee were supported by a Grant from
Korea Research Institute for Standards and Science (KRISS). J. W.
Lee was supported by the Ministry of Science and Technology of
Korea. J. Bergou is grateful for the hospitality and financial
support extended to him during the $2^{nd}$ KIAS-KAIST Workshop in
Seoul.
\end{acknowledgments}


\begin{thebibliography}{99}

\bibitem{BB84}C. H. Bennett and G. Brassard: in Proceedings of
the IEEE International Conference on Computers, Systems and Signal Processing, Bangalore, India (IEEE, New York, 1984), 1996.

\bibitem{Bennett}C. H. Bennett, F. Bessette, G. Brassard, L. Salvail, and J. Smolin: J. Cryptology {\bf 5}, 3 (1992).

\bibitem{Gisin}N. Gisin, G. Riborty, W. Tittel, and H. Zbinden: Rev. Mod. Phys. {\bf 74}, 145 (2002).

\bibitem{Lund}A. P. Lund and T.C. Ralph: Phys. Rev. A {\bf 66}, 032307 (2002).

\bibitem{Pegg}D.T. Pegg, L.S. Phillips, and S.M. Barnett: Phys. Rev. Lett. {\bf 81}, 1604 (1998).

\bibitem{Dakna}M. Dakna, J. Clausen, L. Kn\"{o}ll, and D.-G. Welsch: Phys. Rev. A {\bf 59}, 1658 (1999).

\bibitem{D'Ariano}G. M. D'Ariano, L. Maccone, M. G. A. Paris, and M. F. Sacchi: Phys. Rev. A {\bf 61}, 053817 (2000).

\bibitem{Paris}M. G. A. Paris: Phys. Rev. A {\bf 62}, 033813 (2000).

\bibitem{Resch}K. J. Resch, J. S. Lundeen, and A. M. Steinberg: Phys. Rev. Lett. {\bf 88}, 113601 (2002).

\bibitem{Lvovsky}A. I. Lvovsky and J. Mlynek: Phys. Rev. Lett. {\bf 88}, 250401 (2002).

\bibitem{Davidovich}L. Davidovich, N. Zagury, M. Brune, J. M. Raimond, and S. Haroche: Phys. Rev. A {\bf 50}, R895 (1994).

\bibitem{Moussa}M. H. Y. Moussa and B. Baseia: Phys. Lett A {\bf 245}, 335 (1998).

\bibitem{Freyberger}M. Freyberger: Phys. Rev. A {\bf 51}, 3347 (1995).

\bibitem{Lee1}H. W. Lee and J. Kim: Phys. Rev. A {\bf 63}, 012305 (2000).

\bibitem{Lee2}H. W. Lee: Phys. Rev. A {\bf 64}, 014302 (2001).

\bibitem{Villas-Boas}C. J. Villas-Boas, N. G. de Almeida, and M. H. Y. Moussa: Phys. Rev. A {\bf 60}, 2759 (1999).

\bibitem{Koniorczyk}M. Koniorczyk, Z. Kurucz, A. G\'{a}bris, and J. Janszky: Phys. Rev. A {\bf 62}, 013802 (2000).

\bibitem{Babichev}S. A. Babichev, J. Ries, and A. I. Lvovsky: Europhys. Lett. {\bf 64}, 1 (2003).

\bibitem{Lombardi}E. Lombardi, F. Sciarrino, S. Popescu, and F. De Martini: Phys. Rev. Lett. {\bf 88}, 070402 (2002).

\bibitem{Giacomini}S. Giacomini, F. Sciarrino, E. Lombardi, and F. De Martini: Phys. Rev. A {\bf 66}, 030302(R) (2002).

\bibitem{Enk}S. J. van Enk: Phys. Rev. A {\bf 72}, 064306 (2005).

\bibitem{Lee3}J. W. Lee, E. K. Lee, Y.W. Chung, H. W. Lee, and J. Kim: Phys. Rev. A{\bf 68}, 012324 (2003).

\bibitem{Giorgi}G. L. Giorgi: Phys. Rev. A 71, 064303 (2005).

\bibitem{B92}C. H. Bennett: Phys. Rev. Lett. {\bf 68}, 3121 (1992).

\bibitem{Shannon}C. E. Shannon: Bell System Tech. J. {\bf 27}, 379
(1948); {\bf 27}, 623 (1948).

\bibitem{Caves}C. M. Caves and P. D. Drummond: Rev. Mod. Phys. {\bf 66}, 481
(1994)

\bibitem{Yuen}H. P. Yuen and M. Ozawa: Phys. Rev. Lett. {\bf 70}, 363 (1993)

\bibitem{Holevo1}A. S. Holevo: Probl. Inf. Transm. {\bf 9}, 177 (1973).

\bibitem{Holevo2}A. S. Holevo: IEEE Trans. Inf. Theory {\bf 44}, 269 (1998).

\bibitem{Cabello}A. Cabello: Phys. Rev. Lett. {\bf 85}, 5635 (2000)

\bibitem{BHH}For a recent review on optimized state discrimination strategies see, for example, J. A. Bergou, U. Herzog, and M. Hillery in Quantum State Estimation,
edited by M. Paris and J. \v{R}eh\'{a}\v{c}ek: Lecture Notes in Physics, Vol. 649
(Springer, Berlin, 2004), pp. 417-465.

\end{thebibliography}
\end{document}